\documentstyle[epsf,eqsecnum,aps,prd]{revtex}

\def\w{{\omega}}
\newcommand{\simgt}{\lower.5ex\hbox{$\; \buildrel > \over \sim \;$}}
\newcommand{\simlt}{\lower.5ex\hbox{$\; \buildrel < \over \sim \;$}}
\def\fr{{f_r}}

\begin{document}
\title{Modulation of a Chirp Gravitational Wave from a Compact Binary
due to Gravitational Lensing}

\vfill
\author{Kazuhiro Yamamoto}
\bigskip
\address{
Department of Physical Science, Hiroshima University,
Higashi-Hiroshima 739-8526,~Japan}


\maketitle
\begin{abstract}
A possible wave effect in the gravitational lensing phenomenon
is discussed. We consider the interference of two coherent 
gravitational waves of slightly different frequencies 
from a compact binary, due to the gravitational lensing 
by a galaxy halo. 
This system shows the modulation of the wave amplitude. 
The lensing probability of such the phenomenon is of order 
$10^{-5}$ for a high-$z$ source, but it may be advantageous 
to the observation due to the magnification of the amplitude.
\end{abstract}
\vspace{0.6cm}


\def\M{{M}}
\def\Re{{R_E}}
\def\Hz{{\rm Hz}}

The wave effect in gravitational lensing phenomenon has been 
investigated by many authors (see e.g., \cite{Deguchi,SEF} and 
references therein).
This subject is recently revisited by several authors, motivated 
by a possible phenomenon which might be observed in the future 
gravitational wave experiments 
\cite{TTN,DN,BHN,PINQ,Ruffa,TN,Takahashi,JPM,TNB,TSM}.
In these works, the authors focus on the diffraction in the 
wave effect, which is substantial for $\lambda\sim\Re$, where 
$\lambda$ is the wave length and $\Re$ is the Schwarzschild 
radius of the lens mass. 
However, in the present work, we focus on another different 
aspect of the wave effect, the modulation of superposed two waves,  
in the limit that the geometrical optics is valid 
$\lambda\ll\Re$. Some aspect of this effect has been 
investigated by the author and Tsunoda \cite{KY} as a
lensing phenomenon by a cosmic string. Here we perform 
the similar analysis for the lensing by a galaxy halo.
Throughout this paper, we use the convention $G=c=1$.

We start by considering the superposition of two lensed 
waves with the amplitude 
$A_1$ and $A_2$ and the angular frequencies $\w_1$ and $\w_2$. 
Namely, we consider the wave expressed by
\begin{eqnarray}
  {\cal E} = A_1\cos\left(\omega_1t\right) +A_2\cos\left(\omega_2t+\delta\right),
\end{eqnarray}
where we assume that the phase $\delta$ involves the information of 
the path difference, and the difference of the frequencies $\w_1$ 
and $\w_2$ are due to the time delay effect.
Assuming $A_2=A_1+\Delta A$ and $\Delta A<< A_1$, $\cal E$ reduces to 
\begin{eqnarray}
  {\cal E} &\simeq&\sqrt{2(A_1^2+A_2^2)}
  \biggl[ \cos\left({(\omega_1+\omega_2)t+\delta\over2}\right)
         \cos\left({(\omega_1-\omega_2)t-\delta\over2}\right) 
\nonumber
\\ && \hspace{0.5cm}     
  + {\Delta A\over 2 A_1}
         \sin\left({(\omega_1+\omega_2)t+\delta\over2}\right)
         \sin\left({(\omega_1-\omega_2)t-\delta\over2}\right)
  \biggr].
\label{emod}
\end{eqnarray}
This means that, if $\Delta A/A_1\ll 1$, the wave amplitude 
modulates with the period 
\begin{eqnarray}
  {\cal T} = {4\pi\over (\omega_1-\omega_2)}
\simeq{4\pi\over \Delta T}
\left( d\omega\over dt\right)^{-1},
\end{eqnarray}
where $\Delta T$ is the time delay and $d\omega/dt$ is the change rate of $\omega$
for an observer.

We consider a gravitational wave from a binary of compact objects with equal mass $M$.
By decreasing energy due to the gravitational wave radiation, 
the orbit of the binary changes.
Thus the angular frequency $\omega$ changes. The rate 
of the change is estimated as \cite{Schutz},
\begin{eqnarray}
  {d\omega \over dt}={0.3\times 10^{-8}} (1+z_s)^{5/3} 
  \biggl({M\over M_\odot}\biggr)^{5/3}
  \biggl({\omega\over {\rm rad/s} }\biggr)^{11/3}
  {\rm rad/s^2},
\end{eqnarray}
where $z_s$ is the redshift of the source. Here note that $\omega$ and 
$t$ are the angular frequency and the time of the observer.
Now we consider a lens halo modeled by the singular isothermal sphere
with the velocity dispersion $\sigma$. Then the time delay is
\begin{eqnarray}
 \Delta T=2.7\times 10^7 (1+z_l) 
  \biggl({\sigma\over 200{\rm km/s}}\biggr)^4 
  \biggl({D_{\rm OL}D_{\rm LS}\over H_0^{-1} D_{OS}}\biggr) 
  \biggl({\fr-1\over \fr+1}\biggr)
 ~{\rm s},
\end{eqnarray}
where $D_{\rm OL}$, $D_{\rm OS}$ and $D_{\rm LS}$ are the angular 
diameter distances following the usual convention \cite{SEF}, 
$\fr$ is the flux ratio of the two waves, $z_l$ is the 
redshift of the lens, and we have adopted the Hubble parameter 
$H_0=70{\rm km/s/Mpc}$.
Then the period of the modulation is estimated as
\begin{eqnarray}
  {\cal T} &\simeq& {3.5\times{10^{6}} \over (1+z_l)(1+z_s)^{5/3}} 
  {\fr+1\over \fr-1}
\nonumber
\\
  &&\hspace{0mm}\times\biggl({M\over M_\odot}\biggr)^{-5/3}
  \biggl({\nu\over 0.01{\rm Hz} }\biggr)^{-11/3}
  \biggl({\sigma\over 200{\rm km/s}}\biggr)^{-4} 
  \biggl({D_{\rm OL}D_{\rm LS}\over H_0^{-1} D_{OS}}\biggr)^{-1}
  ~{\rm s}.
\end{eqnarray}

Now let us consider the probability of such a lensing phenomenon.
The optical depth is estimated by \cite{SEF}
\begin{eqnarray}
 \tau(z_s)=\int_0^{z_s}dz_l {dt\over dz_l} \int dM \pi a^2(\sigma[M],z_l,z_s) (1+z_l)^3 
  {dn(M,z_l)\over dM},
\label{tautau}
\end{eqnarray}
where $\pi a^2(\sigma,z_l,z_s)$ is the cross section of the lensing
event for a halo and $dn/dM$ is the mass function. 
We have computed the optical depth by the numerical integration,
adopting the analytic method for the mass function
based on the Press-Schechter formalism \cite{NS,TC}.
Here we assumed the concordance cosmological model: The flat 
universe with the matter density parameter $\Omega_m=0.3$ 
and the Harrison Zeldovich spectrum $n=1$ normalized as $\sigma_8=0.9$.

The wave amplitudes $A_1$ and $A_2$ must be almost the same for the 
modulation formula (\ref{emod}), 
which is satisfied when the flux ratio $\fr$ is nearly equal 1.
Figure 1 demonstrates the optical depth $\tau$ as the function of the
source redshift $z_s$. The upper and lower curves 
assume $\fr<1.4$ and $\fr<1.2$, respectively. 
In the computation of $\tau$, we set the condition 
that the time delay $\Delta T$ is less than $10^8$ s.
It is clear that the optical depth is typically of order 
$10^{-6}\sim10^{-4}$ for a source at redshift 
$z_s\simgt 1$. 

The sum of the amplification factor of two images is written 
as $\mu_1+\mu_2=2(f_r+1)/(f_r-1)$ with the flux ratio $f_r$
for the singular isothermal sphere model \cite{SEF}. 
The energy flux is in proportion to the square of the amplitude 
of the gravitational wave. Therefore the amplification of the 
gravitational wave amplitude due to the gravitational lensing 
can be written as $\sqrt{2(f_r+1)/(f_r-1)}$. If the flux ratio
is very close to $1$, the magnification factor can be large.
It might allow us to observe the gravitational sources at very 
high redshift. 

In conclusion, when a detector with sufficient angular resolution and 
sensitivity is assumed, the periodic modulation in the gravitational wave 
can be a signature of the interference by the lensing 
phenomenon. 
The probability of the lens phenomenon is small, being typically 
of order $10^{-6}\sim10^{-5}$ ($10^{-5}\sim10^{-4}$) for a 
source at redshift $z_s\sim 1 (\sim5)$.
However, the flux of such a lensed source is amplified by the 
magnification effect. 
Therefore such a lensed source might be advantageous to be observed 
in the future gravitational wave experiments (see e.g., 
\cite{SKN} and references therein).

In the present work, however, an idealized situation is assumed.
We assumed a point source, which is a relevant assumption 
for the compact binary as a gravitational wave source. 
It might not be realistic to model the lensing halos with
the spherical symmetry, however, we believe 
that this will not affect our conclusion qualitatively
because we worked under the condition that the usual 
geometrical optics is valid.

\vspace{1mm}
{\it Acknowledgments} 
The author thanks Y. Kojima for reading the manuscript and 
useful comments. He also thanks R. Takahashi for useful
discussions. He is also grateful to the referee for useful
comments, which helped improve the manuscript. 
This work is supported in part by Grant-in-Aid for
Scientific research of Japanese Ministry of Education, Culture, Sports, 
Science and Technology, No.15740155.

\begin{figure}[b]
\begin{center}
    \leavevmode
    \epsfxsize=12cm
    \epsfbox[20 150 600 720]{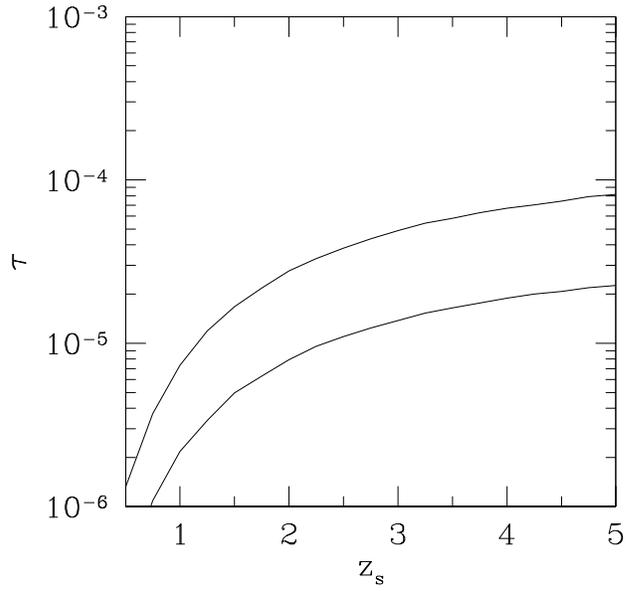}
\end{center}
\caption{Optical depth $\tau$ 
as the function of the redshift of the source $z_s$.
The upper and lower curves correspond to 
$\fr\leq 1.4$ and $\fr\leq 1.2$, respectively. 
We integrated $\tau$ under the condition $\Delta T\leq 10^8 ~{\rm sec}$.}
\label{fig1}
\end{figure}

\end{document}